# A data mining approach for improved interpretation of ERT inverted sections using the DBSCAN clustering algorithm

Kawtar SABOR[1,2], Damien JOUGNOT[1], Roger GUERIN[1], Barthélémy STECK[2], Jean-Marie HENAULT[2], Louis APFFEL[2], Denis VAUTRIN[2]

[1] Sorbonne Université, CNRS, EPHE, UMR 7619 METIS, 75005 Paris, France
[2] EDF-R&D Lab Chatou, France

**Corresponding author**: kawtar.sabor@sorbonne-universite.fr

**Summary:**

Geophysical imaging using the inversion procedure is a powerful tool for the exploration of the Earth's subsurface. However, the interpretation of inverted images can sometimes be difficult, due to the inherent limitations of existing inversion algorithms, which produce smoothed sections. In order to improve and automate the processing and interpretation of inverted geophysical models, we propose an approach inspired from data mining. We selected an algorithm known as DBSCAN (Density-Based Spatial Clustering of Applications with Noise) to perform clustering of inverted geophysical sections. The methodology relies on the automatic sorting and clustering of data. DBSCAN detects clusters in the inverted electrical resistivity values, with no prior knowledge of the number of clusters. This algorithm has the advantage of being defined by only two parameters: the neighbourhood of a point in the data space, and the minimum number of data points in this neighbourhood. We propose an objective procedure for the determination of these two parameters. The proof of concept described here is applied to simulated ERT (Electrical Resistivity Tomography) sections, for the following three cases: two layers with a step, two layers with a rebound, and two layers with an anomaly embedded in the upper layer. To validate this approach, sensitivity studies were carried out on both of the above parameters, as well as to assess the influence of noise on the algorithm's performance. Finally, this methodology was tested on real field data. DBSCAN detects clusters in the inverted electrical resistivity models, and the former are then associated with various types of earth materials, thus allowing the structure of the prospected area to be determined. The proposed data-mining algorithm is shown to be effective, and to improve the interpretation of the inverted ERT sections. This new approach has considerable potential, as it can be applied to any geophysical data represented in the form of sections or maps.

# 1    Introduction

Geophysics is a discipline, the aim of which is to derive information related to geological materials, based on the measurement of physical properties (e.g. the measurement of electrical potential can be used to infer the material's electrical resistivity; the propagation time of mechanical waves can be used to retrieve its associated wave velocities). Geophysical measurements are widely used because they make it possible to use non-destructive techniques to obtain an extended spatial view of the prospected area, as opposed to other destructive techniques used for geotechnical measurements, which produce local information only. An inversion operation needs to be applied to the measured data, in order to determine a physical model of the prospected area, which is able to explain the measurements in terms of known geophysical parameters (e.g. Menke, 1989). Geophysical measurements can be used for various applications. Examples would include applications in geological prospection, based on electromagnetic and seismic measurements (e.g. Bauer et al., 2010; Finco et al., 2018; Hsu et al., 2010), earthen embankment diagnosis and monitoring using electrical measurements to detect possible leakages or weakened areas (e.g. Bièvre et al., 2017; Fargier et al., 2014; Johansson & Dahlin, 1996; Ling et al., 2019), archaeological prospection by means of electromagnetic sounding (e.g. Simon et al., 2012 ; Thiesson et al., 2011), water table monitoring using seismic measurements (e.g. Garambois et al., 2019; Goldman et al., 1989; Pasquet et al., 2015), and landslide characterization using seismic refraction tomography (e.g. Samyn et al., 2012; Uhlemann et al., 2016).

Improved techniques for the processing and interpretation of geophysical data are constantly gaining the attention of various research teams. The inversion of geophysical data often requires the use of constraints, in order to address the problems of under-determination and equivalence, which are inherent to its physics. The frequently encountered smoothed, inverted sections, which are obtained as a result of the regularization process (e.g., Tarantola, 2005), make it difficult to extract accurate information related to the prospected subsoil structures, such as interface positions, and the extent and shape of anomalies. The interpretation of inverted sections is also biased by the colour scale used to represent them (Borland & Russell, 2007; Nicollo, 2014).

Conversely, emerging automatic data analysis techniques based on data mining algorithms have stimulated their application in different fields. In particular, data mining is already used to automatically explore, classify and synthetize texts in the frame of text mining (Tan, 1999; Amado et al., 2018; Westergaard et al., 2018). It is also used in the medical field, for instance to analyze patients' medical and genetic information for the prediction of heart disease (Soni & Ansari, 2011). In geophysics, previous studies have used neural networks: to improve the inversion process (e.g. Krasnopolsky & Schiller, 2003; Russell, 2019; Zheng et al., 2019; Jin et al., 2019), to automatically pick seismic first arrivals (e.g. Chen, 2017), to interpret ERT time-lapse measurements (e.g. Xu et al., 2017), and for parameter estimation (e.g. Calderon-Macias et al., 2000).

In the present study, we propose a proof of concept regarding the use of a specific data-mining algorithm to better interpret inverted geophysical sections. Many studies have dealt with discontinuities and interface characterization, using different methods: analyzing the electrical resistivity gradient and assigning the interface to the steepest gradient area (Chambers et al., 2014), conducting a probabilistic inversion with interface reconstruction and updating, at each step of the inversion (De Pasquale et al., 2019), and performing fuzzy clustering of the inverted geophysical section (Ward et al., 2014). These methods require prior knowledge of the geological structure. The proposed approach enables the automatic determination of shapes and interfaces, leading to a more accurate interpretation of the geophysical data, without any prior knowledge, unlike other approaches (e.g., Dezert et al., 2019). To illustrate this new approach, we selected data recorded using the 2D ERT (Electrical Resistivity Tomography) technique, because it is one of the most commonly used geophysical methods. ERT measurements also provide a very good example of smooth imagery. Indeed, inverted resistivity sections are derived from an integrative physical method, based on diffusive processes (e.g., see discussion in Jougnot et al., 2018 ), which is enhanced by the regularization of the inversion process (e.g., Day-Lewis et al., 2005). As it can be very challenging to interpret the resulting, smooth sections, this represents a perfect field of application for our proposed algorithm.

We first introduce ERT measurements, and describe the processes used by the data mining technique, in particular for the case of ERT data where it is applied to the automatically interpret geophysical sections. We then present simulated sets of 2D ERT data as well as real field data, which are processed according to the proposed methodology. Finally we discuss the sensitivity of the algorithm to its parameters and to noise.

## 2 Materials and methods
### 2.1 Selected geophysical method: Electrical resistivity tomography

Electrical resistivity measurements involve the injection of a DC (Direct Current) electric current into the ground, using two electrodes, and measuring the resulting potential with another two electrodes (Keller & Frischknecht, 1966). Measurements can be recorded using 1D, 2D or 3D electrode configurations (Dahlin, 2001). 1D configurations are used for profiling or electrical sounding. Profiling allows lateral variations in resistivity to be analyzed at a fixed depth, while electrical sounding allows the vertical variations in resistivity to be analyzed at a fixed lateral position. 2D and 3D configurations are a combination of both profiling and electrical sounding, which allows 2D resistivity maps or 3D resistivity tomograms to be retrieved. 2D and 3D measurements are called ERT. For a more detailed description of 1D, 2D and 3D configurations, see Binley (2015). In the present study, 2D electrical resistivity tomography measurements are considered because they correspond to the most commonly used ERT application. The relationship between the source location $\mathbf{r_s}$, its intensity $I$ and the resulting electrical potential field $\mathbf{V}$ is described by combined Maxwell's equations (Kunetz, 1966):

$$\boldsymbol{\nabla}.\left(\frac{\mathbf{1}}{\boldsymbol{\rho}}\boldsymbol{\nabla}\mathbf{V}\right) = -I\delta(\mathbf{r} - \mathbf{r_s}), \tag{1}$$

with $\mathbf{r_s}$ being the position of the injecting source, $\mathbf{r}$ the measurement position, $I$ the injected current, $\boldsymbol{\rho}$ the electrical resistivity and $\mathbf{V}$ the electric potential.

We denote by $\boldsymbol{V}_{\text{mes}}$ the vector containing the values obtained with a given measurement protocol. The aim of the inversion process is to retrieve the underground electrical resistivity distribution from the potential differences $\boldsymbol{V}_{\text{mes}}$ measured at the surface. For this, a forward model is used to generate simulated values of electric potential, based on knowledge of the underground electrical resistivity distribution and the location and intensity of injected currents. In general, a finite-elements method is used to numerically solve the combined Maxwell's equations. In the following, we denote by $\boldsymbol{\rho}$ the vector containing the discretized electrical resistivity values, and by $f$ the forward model function.

Due to the ill-posed problem of the inversion (Ellis & Oldenburg, 1994), $\boldsymbol{\rho}$ cannot be derived directly from the measurements $\boldsymbol{V}_{\text{mes}}$. This problem is generally solved by minimizing a regularized least-squares criterion (Rücker et al., 2006):

$$\phi(\boldsymbol{\rho}) = \phi_{\text{d}}(\boldsymbol{\rho}) + \phi_{\text{r}}(\boldsymbol{\rho}), \tag{2}$$

where $\phi_d$ is a data misfit term corresponding to the sum of the squared differences between the measurements and the output produced by the forward model:

$$\phi_d(\boldsymbol{\rho}) = \|\mathbf{W}(\boldsymbol{V}_{\text{mes}} - f(\boldsymbol{\rho}))\|_2^2, \quad (3)$$

where $\mathbf{W}$ is a weighting matrix derived from the measurement uncertainties, estimated during the acquisition procedure, and $\phi_r(\boldsymbol{\rho})$ is a regularization term which allows prior information related to the underground characteristics to be included. The first-order Tikhonov regularization scheme is very commonly used in the framework of geophysical tomography (Cardarelli & Fischanger, 2006; Ditmar & Makris, 1996). This method relies on the penalization of spatial variations in the estimated electrical resistivity field, in both the horizontal and vertical directions:

$$\phi_r(\boldsymbol{\rho}) = \lambda_x \|\mathbf{D}_x \boldsymbol{\rho}\|_2^2 + \lambda_z \|\mathbf{D}_z \boldsymbol{\rho}\|_2^2. \quad (4)$$

The matrices $\mathbf{D}_x$ and $\mathbf{D}_z$ are used to compute the first-order spatial derivatives along the $x$ and $z$ axes, and $\lambda_x$ and $\lambda_z$ are two regularization parameters, which balance the impact of each member in the minimized criterion. This regularization technique enforces the reconstruction of smooth areas, which is consistent with the fact that the underground medium is composed of several homogeneous layers of unknown shape. However, this regularization scheme tends to make real sharp resistivity variations appear to be excessively smooth, such that the boundaries between layers cannot be accurately identified in the reconstructed underground image (e.g., Day Lewis et al., 2005).

For the present study, we used the open-source pyGIMLI package for inversion (python Geophysical Inversion and Modelling Library, Rücker et al., 2017). This package, which makes it possible to model and invert ERT measurement data as described above, was chosen for its ability to rapidly and straightforwardly create different underground geometries, and to freely set the inversion parameters. A further advantage of this package is that its library can be used with other geophysical methods (e.g., induced polarization and seismic refraction).

**2.2 Description of the data mining approach**

Data mining refers to the process of analyzing large data sets to extract information for further use (e.g. decision-making, trends, patterns or class determination). This process involves several steps, which depend on the field of application, and can be transposed to the interpretation of geophysical data (Han et al., 2011). In the present study, we adapted the procedure to the case of ERT (

Table *1*). Note that $\rho_{inv}$ refers to the inverted electrical resistivities. In the following, the term "data" generally refers to different variables, depending on the step being applied (

Table *1*), whereas the "data" referred to in "data mining" and the term "data point" refer to the logarithm of $\rho_{inv}$.

**Table 1** Data mining steps in the particular case of ERT data

| Data mining step | Corresponding step in the geophysical application to ERT |
|---|---|
| a. Data selection | Select the geophysical measurement, which in the case of our study is the electric potential, as well as the measurement configuration and parameters needed to obtain the apparent electrical resistivity |
| b. Data pre-processing | Invert the measured values of electric potential and represent the |
|   | profiles in tabular form, using the parameters x, z, $\rho_{inv}$ |
| c. Data transformation | Compute the logarithm of each pre-processed data point $\rho_{inv}$ |
| d. Data mining | Apply the clustering algorithm: DBSCAN (Density-Based Spatial Clustering of Applications with Noise, Ester et al., 1996) to the data produced by step (c) |
| e. Interpretation | Identify the geological structure corresponding to the inverted data. |

**a. Data selection**: This is an important step in the general use of data mining and in particular for the case of geophysical applications. It consists in choosing the parameter that could contain information related to what we are seeking. In the present case, the appropriate geophysical method is chosen for its sensitivity to the properties of interest, which depend on: the aim of the prospection (e.g. geological structure characterization, leakage monitoring, anomaly detection), the nature of the prospected area (e.g. silty soil, clay-rich soil) and the desired spatial resolution. These criteria are also taken into account when selecting the measurement parameters (e.g. number of probes, probe spacing, type and amplitude of the sources). In the present case, we consider ERT measurements only, although a similar approach could be used with any kind of geophysical profile or map.

**b. Data pre-processing:** This step consists in cleaning the raw measurements, removing invalid measurement points, and applying any necessary calculations (e.g., if the raw measurements do not contain information of interest for the study). In the present case, invalid measurement points that are attributable, for example, to poor coupling of the probes must be removed prior to the inversion process used to retrieve the electrical resistivity $\rho_{inv}$. In order to simplify manipulation of the pre-processed measurements, and to apply suitable algorithms in the following steps, the data needs to be represented in a standard format, i.e. typically tables and matrices. In the present case, our inverted ERT profiles are converted to a listing in the form of a table: *x* (horizontal position), *z* (vertical position) and $\rho_{inv}$, where *x* and *z* are the coordinates of the centres of the cells in the inversion mesh.

**c. Data transformation:** In order to enhance the contrast in the data, or to facilitate the extraction of information, transformations can be applied (e.g., the square of the data, the logarithm of the data). In the present case, as we usually analyse the logarithm of the resistivity in the context of ERT interpretations, a logarithmic operation was applied to the inverted electrical resistivity.

**d. Data mining**: This is the main step of the proposed approach, as it deals with the data-mining algorithm *per se*. Once the data has been pre-processed and transformed, it is analysed by applying the appropriate algorithms. Three broad classes of algorithm can be distinguished:

(i) Regression algorithms, which try to establish a linear or nonlinear model that is well adjusted to the data.

(ii) Classification algorithms, which consist in allocating data points to different classes, on the basis of their value. Classes could for example include soil types, vegetation types, or a client's gender. Classification is performed in two successive steps. Firstly, data associated with known classes or outputs are fed to the algorithm, which then learns the limits between existing classes or patterns in the data provided. Then, when new data is introduced, the algorithm is able to predict the output or the class, on the basis of what it has learned. It should be noted that classification algorithms require data with known outputs, in order to predict the output of other new data.

(iii) Finally, clustering algorithms are designed to infer clusters or categories present in the data. However, unlike a classification algorithm, no learning step is performed because the clusters are assumed to be unknown. This type of algorithm analyses the data and proposes a set of clusters or patterns in the data. In the case of the present study, as is generally the case, the different soil types or geological composition of the prospected area are unknown.

A clustering algorithm was selected for the research presented here. Note that the position of the data points in the inversion mesh is not included in this algorithm.

**e. Interpretation:** This step uses the results of the previous steps to interpret the data, depending on the aim of the study or analysis. This could, for example, correspond to decision-making, based on a prediction using a model obtained by regression, or a prediction using classification results. It could also involve pattern extraction, based on the results of clustering.

In the present application, as geophysical methods are often used to determine the structure of the prospected areas, clusters could be identified on the basis of electrical resistivity values. These clusters were associated with the different soil types present in the prospected area, which contributed to the assessment of the structure of the prospected area. The same electrical resistivity values were assigned to the same cluster, even when they were associated with different spatial locations.

### 2.3 Chosen data-mining algorithm: DBSCAN

There are many well-known clustering algorithms (e.g., Nagpal et al., 2013; Shirkhorshidi et al., 2014), which differ mainly in terms of the data analysis applied (statistical analysis, density-based analysis) and the data distribution they can handle. Some of these would not be appropriate for our application, as they require a high number of parameters to be set, or they are more specifically designed for Gaussian distributions.

Following a careful review of various existing algorithms (Nagpal et al., 2013; Shirkhorshidi et al., 2014), we selected the so-called DBSCAN algorithm (Density-Based Spatial Clustering of Applications with Noise, proposed by Ester et al., (1996). This algorithm relies on the local analysis of data-point densities in the data space, and forms clusters in an iterative manner. This approach makes it possible to handle data distributions with a wide variety of different shapes (profiles, sections, maps, 3D volume). In addition, only two parameters need to be determined by the user:

- $\varepsilon$ which corresponds to a variable range around a data point. The $\varepsilon$-neighbourhood of a data point P is defined as the interval centred around P, of width equal to $\varepsilon$. This width is defined in terms of the clustering variable (the logarithm of the inverted electrical resistivity in the present case). The spatial coordinates ($x,y$) refer to the horizontal distance and depth of each point, respectively, but are not included in the clustering algorithm. This information is used only to reposition the clustered points at their spatial locations, once the clustering has been performed for all points in the dataset.
- $N$ which corresponds to the required minimum number of data points in the $\varepsilon$-neighbourhood of a point P, for P to be considered as belonging to a given cluster

Figure 1 summarizes how the method is used by the algorithm to analyze each data point, and to assign it to a cluster or to noise. For a data point P, the algorithm counts the number of points $e$ in the $\varepsilon$-neighbourhood of P. If this number of points is greater than or equal to $N$, two possible cases are distinguished. If one of the points in the $\varepsilon$-neighbourhood of P is already assigned to a cluster, P is assigned to that same cluster.

Otherwise, P is assigned to a new cluster and the noise points in the ε-neighbourhood of P are assigned to the same new cluster as P. If *e* is less than *N*, then the points found within the ε-neighbourhood of P are checked. If any of these points have been assigned to a cluster, then P is assigned to the same cluster, otherwise it is considered to be a noise point.

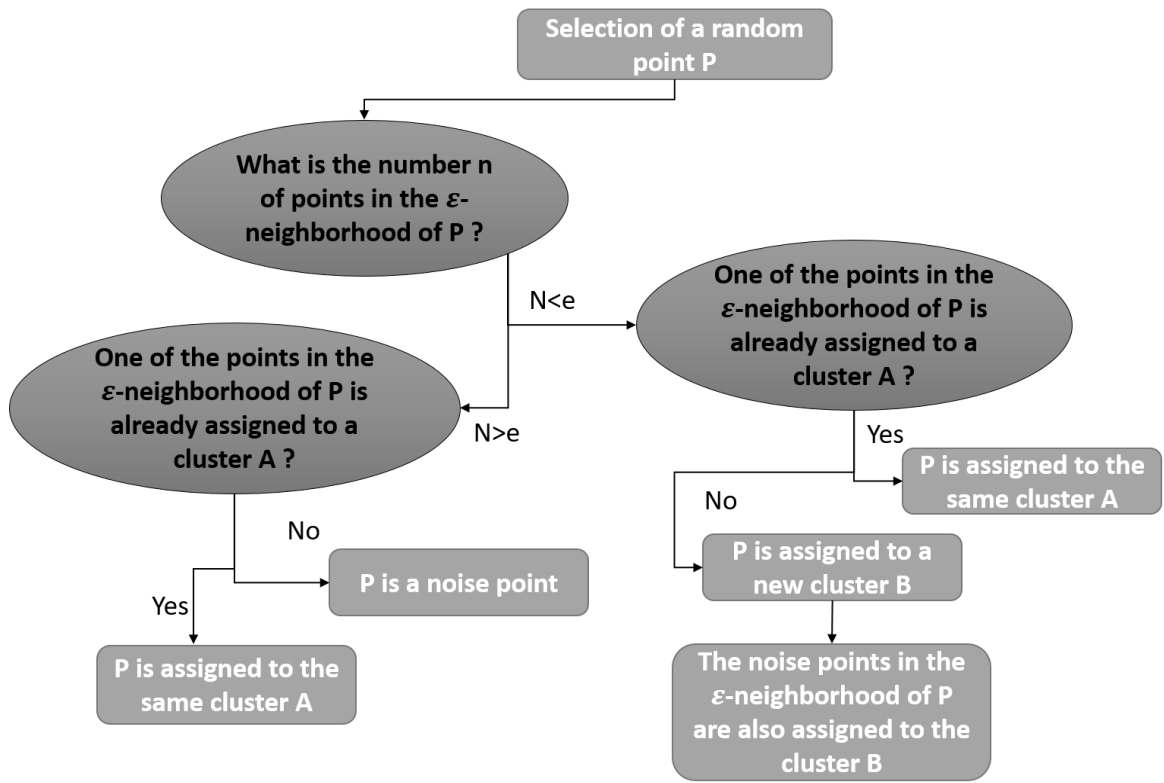

**Figure 1** Flow diagram describing the DBSCAN algorithm and data point analysis methodology

Figure 2 describes the main types of point analysis handled by the algorithm, in the case of electrical resistivity values for a random value of epsilon ε and for *N*=3. For a set of data points, such as that represented in Figure 2, along an axis representing the clustering variable (i.e., the logarithm of the inverted electrical resistivity in the present case), a random point P is chosen and the number of points *e* in the ε-neighbourhood of P is computed, as shown in step (1) of Figure 2. For this first case, *e* is equal to 1 and is then less than the minimum number of points (*N*) required to assign P to a cluster. P is thus considered to be a noise point. In the next step (2), a random second point is chosen, leading to a second noise point. In step (3), the next point Q is explored. As it has four points in its ε-neighbourhood, it is classed as belonging to a cluster. Since all the points in the ε-neighbourhood of Q are either noise points or have not yet been analyzed, a new cluster is created and Q is assigned to that cluster. The noise points found in the ε-neighbourhood of Q are then re-assigned to that cluster. Another case can arise when the selected point has more than *N* points in its ε-neighbourhood, but none of those points belong to any previous cluster. Another new cluster is then defined, as shown in step (4). This analysis is maintained until all the points have been analyzed, as shown in step (5). Then, using the known positions (*x,z*) of each point, the profile is reconstructed and each cluster is associated with a specific soil type.

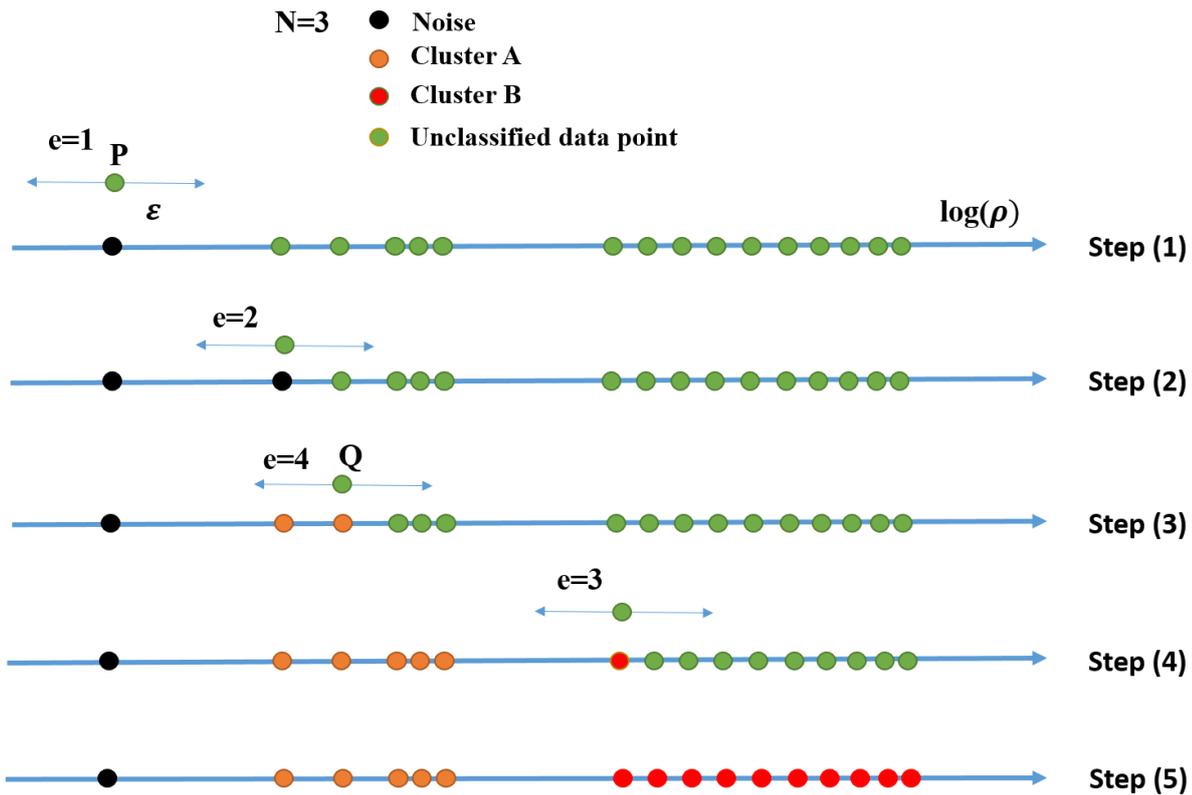

**Figure 2** Illustration of the DBSCAN steps followed in the case of ERT data analysis, for *N* = 3 and a random range of *ε*. These 5 steps summarize different point cases that the algorithm can handle.

Note that the example shown in Figure 2 is highly simplified, since the clusters can be clearly distinguished, with no need for any particular analysis. However, in real cases the data points (i.e., the logarithm of the inverted electrical resistivity values in the present study) are more densely packed, and the boundaries between clusters cannot be distinguished. As an example, Figure 3a shows the electrical resistivity data distribution for a two-layer tabular model. By applying the DBSCAN algorithm to this dataset, with appropriate values of *ε* and *N*, two clusters are clearly distinguished (Figure 3b).

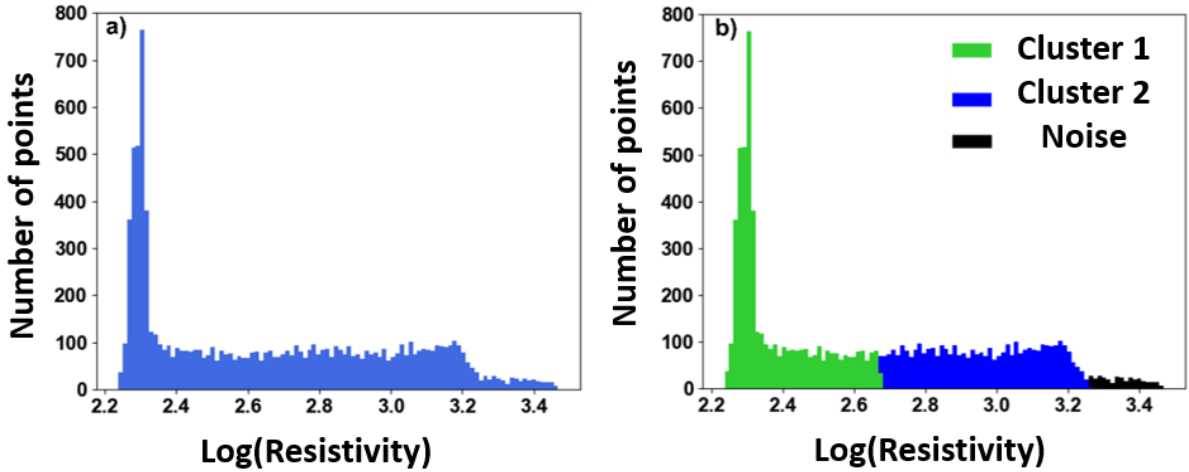

**Figure 3** (a) Example of the distribution of the logarithm of electrical resistivity values, for a two-layer tabular model. (b) Visualisation of the clustering determined when DBSCAN is applied to the resistivity distribution in (a)

The remaining question is the choice of DBSCAN parameters, as these can vary from one DBSCAN field of application to another, as described in the following section.

### 2.4 Choice of DBSCAN parameters

**Choice of *N***

In the case of the present study (i.e., 2D ERT profiles), the numbers of points, *N*, needs to be chosen with respect to the objectives of the ERT prospection campaign, and two special cases need to be distinguished. If the aim is to find anomalies or special objects, *N* should be chosen to be greater than or equal to the expected number of data points needed to cover the anomaly, as defined by the resolution of the inversion mesh. Selecting a higher value of *N* will cause the anomaly to merge with the cluster representing the area containing the anomaly. The second case arises when an accurate description of the general soil structure is needed, with no specific expectation of an anomaly. Under these conditions, we recommend using: $N = \frac{N_{mesh}}{2*k}$ , where $N_{mesh}$ is the number of the data points, $\log(\rho_{inv})$, which is equal to the number of inversion mesh cell centres and *k* is the maximum number of expected soil types. Although the use of a smaller value has no influence on the outcome of the clustering step, the selection of a higher value of *N* can cause certain clusters to merge, and hence mask the presence of some structures. The choice of a very small value of *N* can lead to the detection of clusters that don't really exist, and which are formed only because *N* is too small. For these reasons, the choice of *N* is an essential step for a successful clustering process. These aspects are analysed in the discussion. The criteria used when selecting the most appropriate value for *N* are summarized in *Table 2*.

**Table 2** Suitable *N* value, depending on the aim of the geological prospection. $N_{mesh}$ is the number of data points, which is equal to the number of inversion mesh cells. *k* is the maximum number of expected soil types.

| Aim of the prospection | Suitable *N* value |
|---|---|
| General structural analysis | $\frac{N_{mesh}}{2*k}$ |
| Anomaly detection | Expected number of mesh cells to cover the anomaly |
| Unknown structure | Expected number of mesh cells to cover a potential anomaly |

**Choice of $\varepsilon$**

In order to choose a suitable value for $\varepsilon$, we represent the data-point density using an *N*-dist plot, where *N* is the number of points, as defined in the previous paragraph. Figure 4 provides a diagrammatic description of the process used to obtain this type of plot. First, for each data point P, we compute the mean value of the distances from this point to all of its *N* nearest neighbours, i.e. the *N* closest points to *P* in terms of *resistivity*, rather than spatial distance. This step is executed for all data points (i.e. all inversion mesh cell centre points to which a resistivity value has been assigned). The computed mean values are then sorted in ascending order and plotted. An example of a *N*-dist plot is provided in Figure 5, where the initial *N*-dist curve is shown in Fig. 5a. The most suitable value for $\varepsilon$ is then defined as the value of *N*-dist corresponding to the point of maximum curvature in the *N*-dist plot. To retrieve this point we compute the first and the second derivatives of *N*-dist (Figure 5b and *5*c). The correct value of $\varepsilon$ is then defined by this point of inflection, where the second derivative of the *N*-dist plot reaches a maximum.

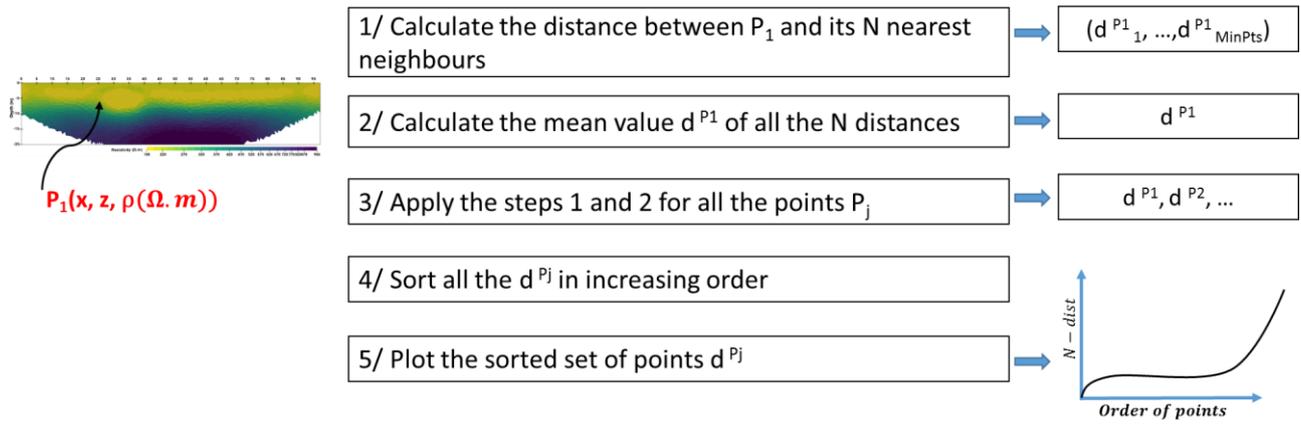

**Figure 4** Diagram describing the steps followed in order to obtain a *N*-dist plot starting from inverted electrical resistivity profile. The mentioned distance refers to the distance between points in parameter space, and is not a spatial distance.

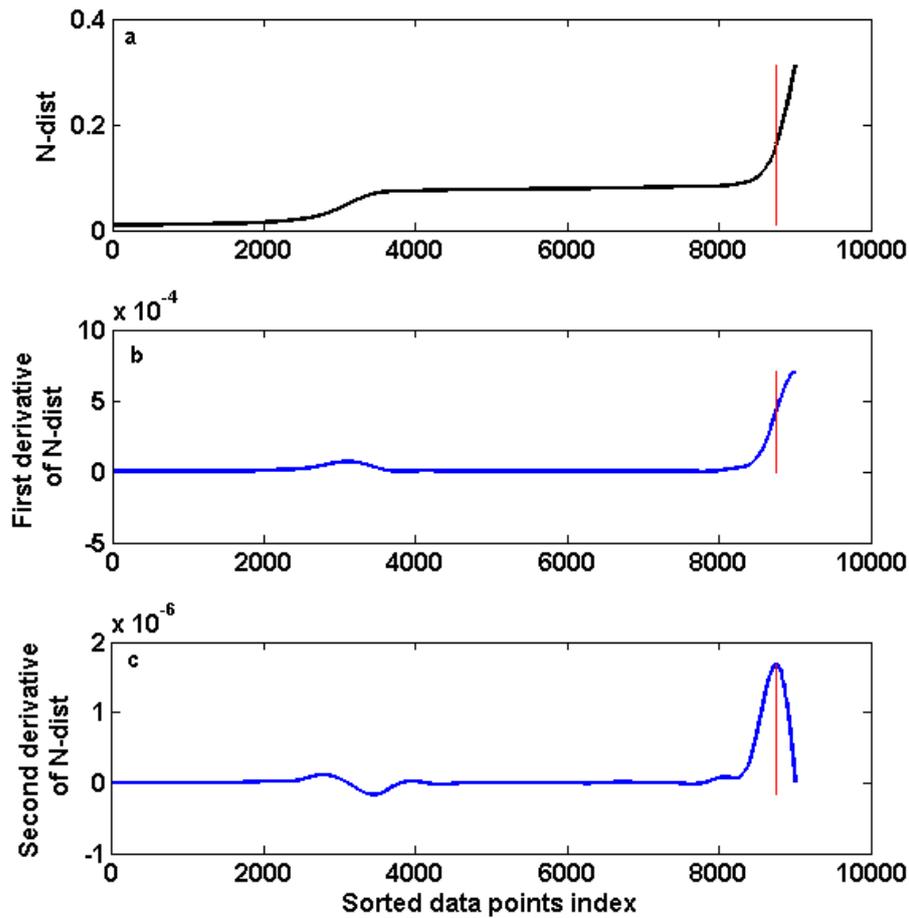

**Figure 5** Example of the analysis of an *N*-dist plot. (a) *N*-dist plot for *N* = 2000. (b) Plot of the first derivative of *N*-dist. (c) Plot of the second derivative of *N*-dist. The vertical red line indicates the position of the maximum curvature point in the *N*-dist plot, corresponding to the maximum value of the second derivative of *N*-dist.

## 3   Simulations

The approach described in the previous section has been applied to three numerical examples, in the context of the present study. These examples are presented in the following, and correspond in all cases to two layers, with (A) a step, (B) a rebound and (C) an anomaly. In the present study, the same electrode configuration is considered: a set of 96 electrodes with 2 m spacings. A Schlumberger reciprocal configuration with 2100 apparent electrical resistivity measurements is considered, allowing a maximum depth of investigation of 32 m to be reached. The inversion was performed using a mesh with *circa* 9000 cells, in order to have a sufficient volume of data for the data-mining algorithm. For the purposes of this analysis, the simulated data are noise free. The aforementioned examples, with a set of 10000 data points, were run in less than 1 minute using an Intel core i7 CPU. This level of computing performance allows this approach to be integrated into a fast workflow for the interpretation of ERT data.

## 3.1 Case A: Two layers with a step

The first example is defined by a tabular model of two layers of soil, with a step in the middle (Figure 6a). The upper layer is conductive, with an electrical resistivity equal to 200 Ωm that could represent clay soil, whereas the lower layer is resistive, with a resistivity equal to 2500 Ωm, corresponding to bedrock. The upper layer has a thickness of 7 m, which then becomes 10 m, corresponding to a 3 m step height. The inversion mesh has 9200 cells. The inverted electrical resistivity profile (Figure 6a) exhibits two structures with a smooth variation, and the exact shape of the interface does not appear clearly. The distribution of the logarithm of electrical resistivity values (Figure 6b) reveals no gaps or clear transitions between the values associated with each of the two layers. The clustered distribution (Figure 6c) confirms that it is not possible to visually detect the difference between the two layers (i.e., between the two clusters) in the raw distribution, because the transition from one cluster representing a specific soil type, to another, does not occur at any specific position and cannot be determined before applying DBSCAN. The clustering is performed using N = 1000 and ε = 0.5. Knowing the position of their associated data points, the two clusters are used to reconstruct the physical makeup of the model (Figure 6d) – it can be seen that its structure is well matched with the originally imposed, two-layer terrain distribution used for the simulation (Figure 6d).

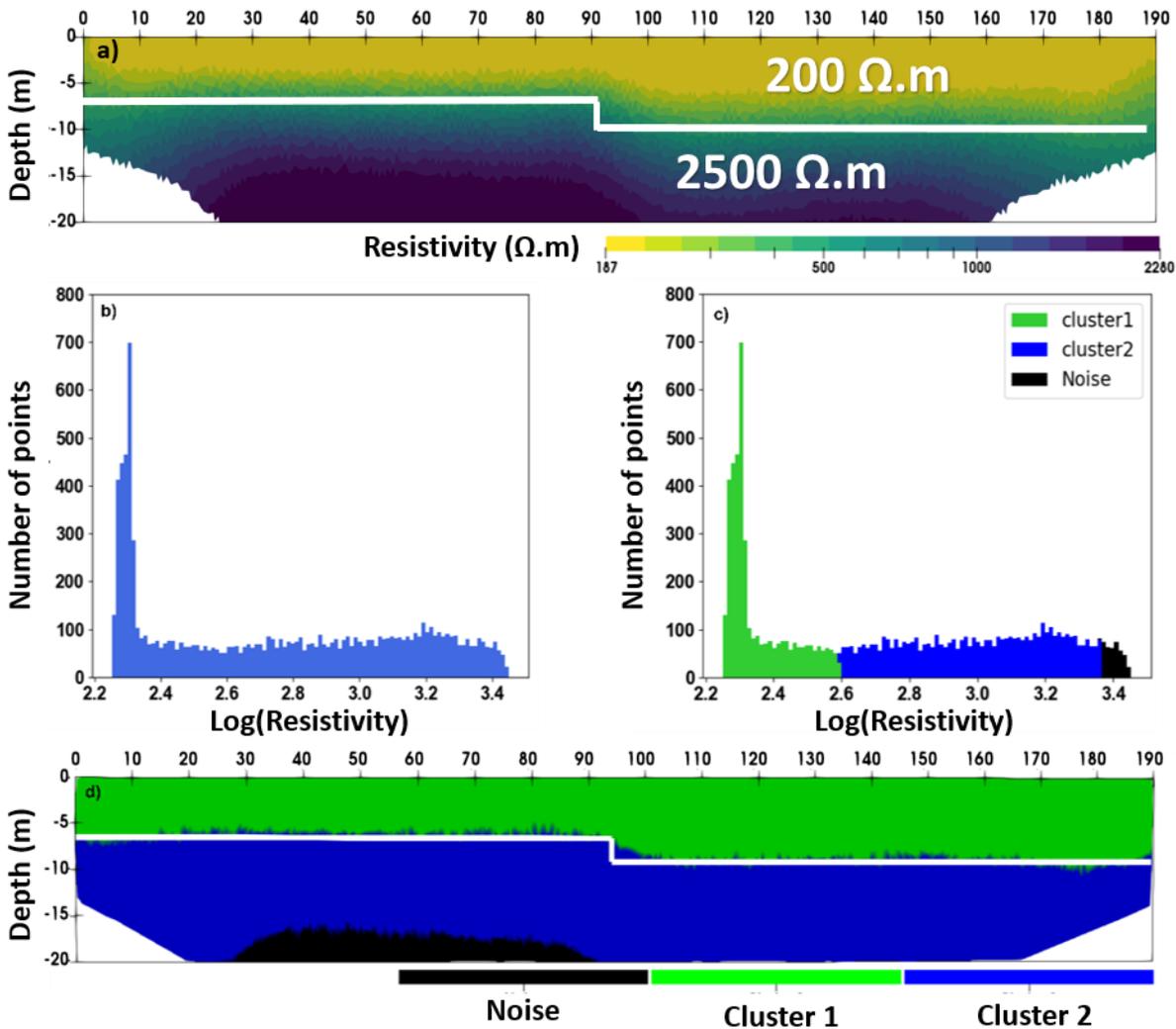

**Figure 6** (a) Inverted electrical resistivity of the tabular two-layer model, characterised by a small step in the middle (case A). (b) Raw distribution of the logarithm of the inverted electrical resistivity values corresponding to case A. (c) Result produced by the DBSCAN clustering algorithm, when applied to the logarithm of the inverted electrical resistivity values in (a). The clustering shown here is implemented using $N$ = 1000 and $\varepsilon$ = 0.5. (d) Result produced by the clustering algorithm, after placing each clustered electrical resistivity value in its correct spatial location. The white line indicates the true position of the interface between the two layers, defined in the model.

### 3.2 Case B: Two layers with a rebound

The second example is defined by a tabular, two-layer model with a rebound in the middle (Figure 7a). The first layer is conductive, with an electrical resistivity equal to 200 Ωm and a thickness of 10 m, whereas the second layer is resistive, with a resistivity equal to 2500 Ωm. The rebound has a height of 6 m. The inversion mesh has 9000 cells. The inverted electrical resistivity profile (Figure 7a) exhibits a blurred structure, and the exact shape of the interface is unclear on the inverted profile. As in the case of the previous example (shown in Figure 3a), the distribution of the logarithm of electrical resistivity values is not characterised by any gaps or clear transitions, from one layer to the other. The clustered distribution (Figure 3b) confirms that it is not possible to visually detect the difference between the two layers in the raw data distribution. When the clustering algorithm was executed using *N* = 2000 and *ε* = 0.16, the retrieved (inverted) structure was found to be very well matched with the initially simulated structure (Figure 7b). Note that the black data points refer to noisy data, i.e. data points with very high inverted resistivity values, which are produced by the influence of rebound effects in the electric field. These noisy data points, which are not considered in the interpretation of the data, are located at depths where the sensitivity of the measurements is not as high as in the shallow sections, especially for the case of the simulated configuration with a conductive layer of soil above a resistive one.

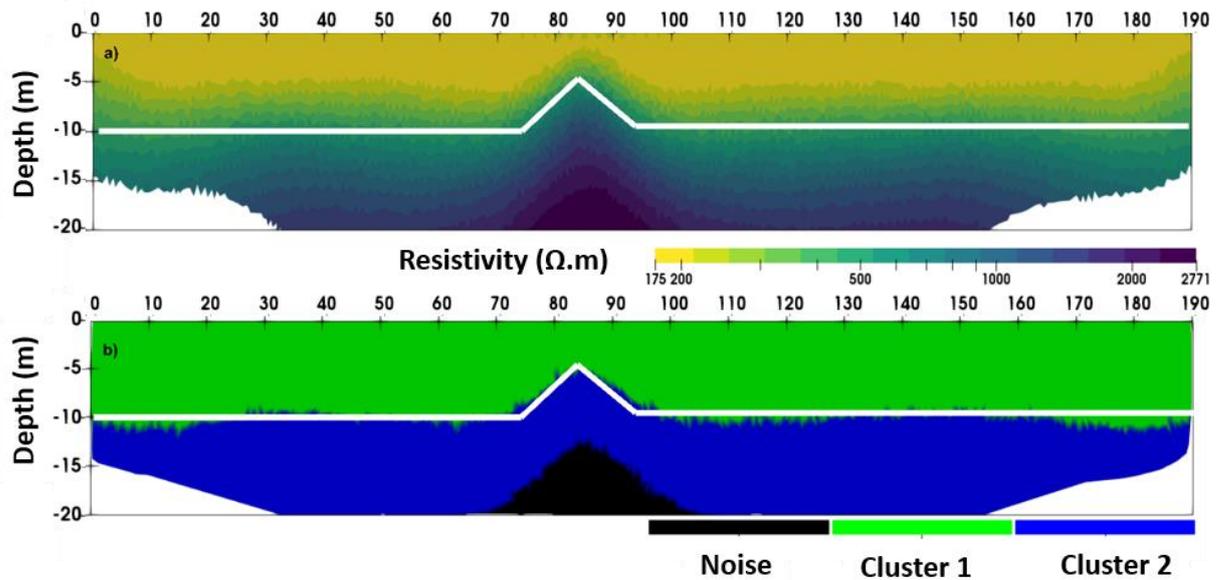

**Figure 7** (a) Inverted electrical resistivity of the tabular two-layer model, characterised by a rebound in the middle, (case B). (b) Result produced by the DBSCAN clustering algorithm, after placing each clustered electrical resistivity value in its correct spatial location. The white line indicates the true position of the interface between the two layers, as defined in the model. The clustering was performed using *N* = 2000 and *ε* = 0.16.

### 3.3 Case C: Two layers and an anomaly

The last example is provided by a tabular model with two layers and a small anomaly embedded in the first layer (Figure 8a). The latter has an electrical resistivity equal to 200 Ωm and a thickness of 10 m, whereas the second layer has an electrical resistivity equal to 2500 Ωm. The anomaly has an electrical resistivity of 25 Ωm and its dimensions are 5 m in length and 2 m in thickness. The inversion mesh has 4500 cells. The inverted electrical resistivity profile (Figure 8a) has two distinct structures with some smooth variations at their interface. Although a vaguely distinguishable, blurred anomaly can be seen, its exact shape and extent are not clear. The same observations can be made as in the case of the previous examples: the DBSCAN algorithm significantly improves the ease with which the structure of the ERT profile can be distinguished, and permits enhanced interpretation of this profile (Figure 8b and 8c and 8d). The clustering was performed using *N* = 500 and *ε* = 0.49. In the retrieved structure, the interface between the two layers is well defined and the anomaly is also detected. Although it is represented by noisy data points, since these are located inside the first layer, which has a good sensitivity, they cannot be neglected in the interpretation of the clustering results, as was the case for the previous rebound example.

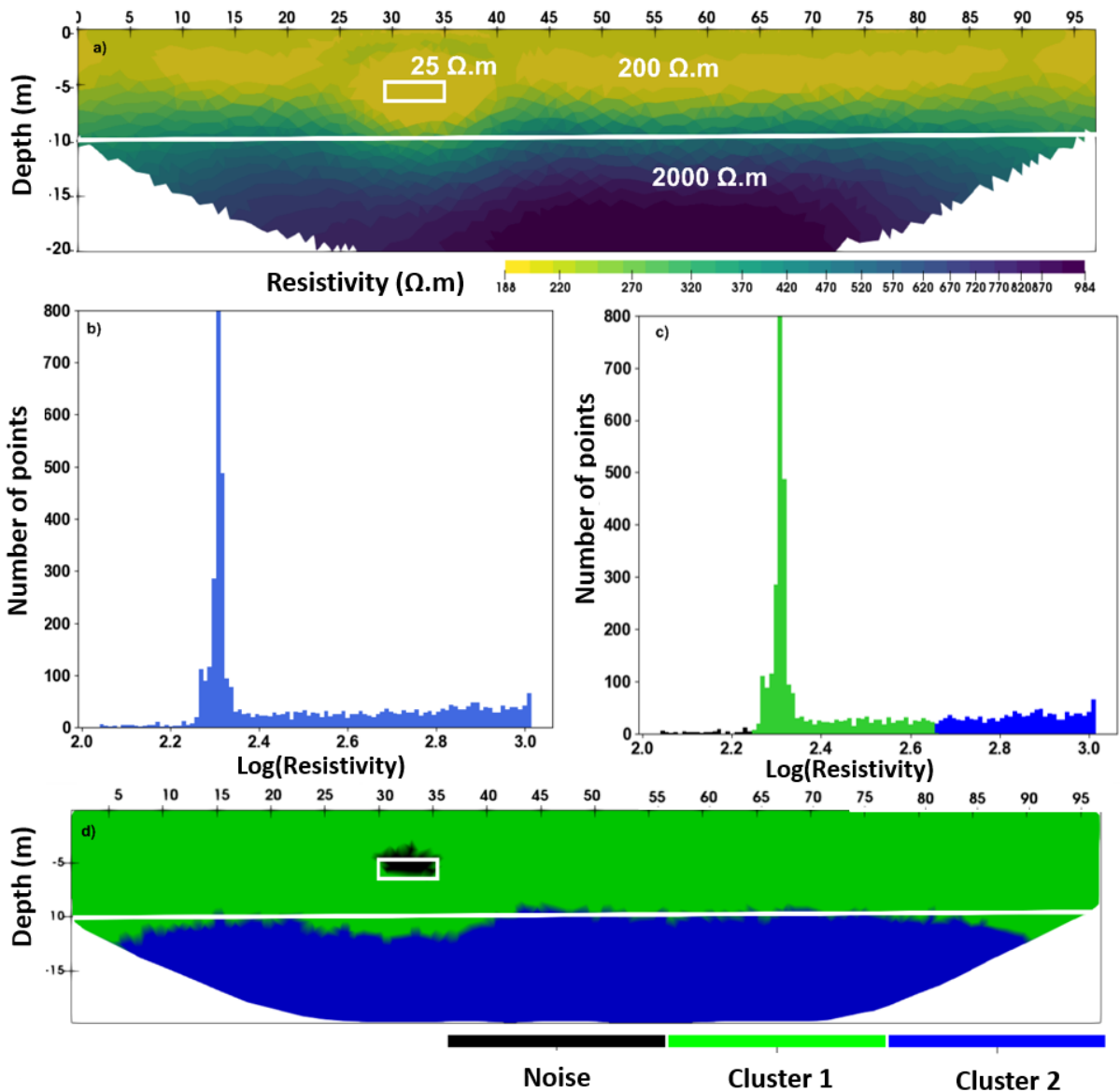

**Figure 8** (a) Inverted electrical resistivity of the tabular two-layer model, characterised by an anomaly, (case C). (b) Raw distribution of the logarithm of the inverted electrical resistivity values corresponding to the two-layer model with an anomaly. (c) Result produced by the DBSCAN clustering algorithm, when applied to the inverted electrical resistivity values in (a). The clustering shown here is implemented using N = 500 and ε = 0.49. (d) Result produced by the clustering algorithm, after placing each clustered electrical resistivity value in its correct spatial location. The white line indicates the true position of the interface between the two layers, as defined in the model.

# 4 Sensitivity analysis

In this section we investigate various properties of the DBSCAN parameters, in an effort to understand how the algorithm works, and how its parameters influence the results.

## 4.1 Analysis of the criterion of the choice of $\varepsilon$

In order to better understand the criteria that should be taken into account when defining the parameter ε, we analyse the *N*-dist plot, for the case of the last example (case C), with *N* = 50. Figure 9 presents the results of the DBSCAN clustering algorithm, through the use of different colours, revealing which of the three different clusters each data point belongs to (Figure 9a). A zoom on the position of the optimal value for $\varepsilon$ (Figure 9b) shows that this position marks the transition between the clustered data points having similar high densities (small distances) and the noisy data points, characterised by a very low density (much greater distances), thus corresponding to outliers or anomalies. In the present example, the first portion of the *N*-dist distribution (bordered by an orange rectangle, Figure 9a) corresponds to the densest set of data points, and comprises the lowest values of *N*-dist. This corresponds to the shallow portion of the structure (Figure 9c), where its inversion is the most accurate and its sensitivity is the highest. As the data points have inverted resistivity values that lie very close to each other, their arrangement is very dense. The second portion of the *N*-dist distribution (bordered by a purple rectangle, Figure 9a) is characterised by a slightly lower density (slightly higher mean distance between points) and corresponds to the deeper part of the structure, which is affected by the strongest smoothing effects produced by the inversion (Figure 9d). Finally, the last portion of the *N*-dist distribution (bordered by a red rectangle, Figure 9a) corresponds to a sparsely populated set of data points, since they represent electrical resistivity values that are very different from those of the surrounding values, and are thus characterised by greater distances. In the present example, these points are related to the anomaly (*Figure 9*e), which has electrical resistivity values that are quite different to those of the surrounding soil in the first layer. This example provides useful insight into the processes implemented by the DBSCAN algorithm, which analyses the data-point density, thus making it possible to distinguish between the different structural components of a prospected area.

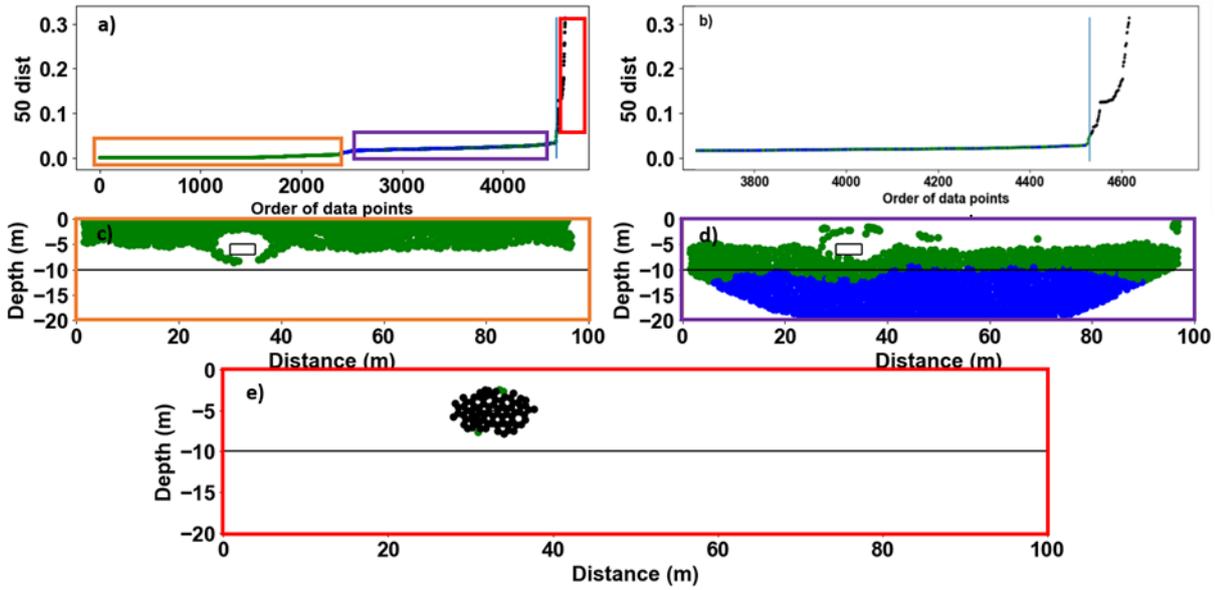

**Figure 9** (a) N-dist plot using different colours to reveal the three different clusters produced by DBSCAN (with N = 50). (b) Zoom on the position of the selected value of epsilon ε located at the interface between the clustered points and the noise data points. (c) Spatial position of the clustered points corresponding to the first portion of the N-dist plot shown in (a), bordered by an orange rectangle. (d) Spatial position of the clustered points corresponding to the second portion of the N-dist plot shown in (a), bordered by a purple rectangle. e) Spatial position of the clustered points corresponding to last portion of the N-dist plot shown in (a), bordered by a red rectangle.

### 4.2 Sensitivity to noise

The previous analysis was applied to simulated models, in the absence of noise. In this section, different noise values are considered in order to assess the impact of noise on the clustering results. For this, a Gaussian noise was added to the simulated values of raw apparent electrical resistivity:

$$\rho_{a\ noise} = \rho_a \left(1 + \mathcal{N}_{0,1} \times error\ (\%)\right), \qquad (5)$$

with $\mathcal{N}_{0,1}$ being a random, centred Gaussian distribution.

The previously described approach was applied to the rebound example, following the addition of respectively 5% and 10% of noise to the raw apparent electrical resistivity data. The resulting distribution of data points, for the three cases of 0%, 5% and 10% noise (Figure 10), is found to lead to smoothing of the (DBSCAN) reconstructed structures, which increases as the noise increases (Figure 11). Although these reconstructed structures are similar to those used in the simulated model, the position of the interface becomes increasingly inaccurate as the noise increases. This effect can be understood as follows: the addition of noise smooths the data distribution, such that the data points tend to have more similar values of inverted electrical resistivity and are thus interpreted to belong to the same cluster. This forces the transition from the first to the second cluster to occur at greater depths than in the original model. The addition of noise also increases the number of data points in the noise cluster.

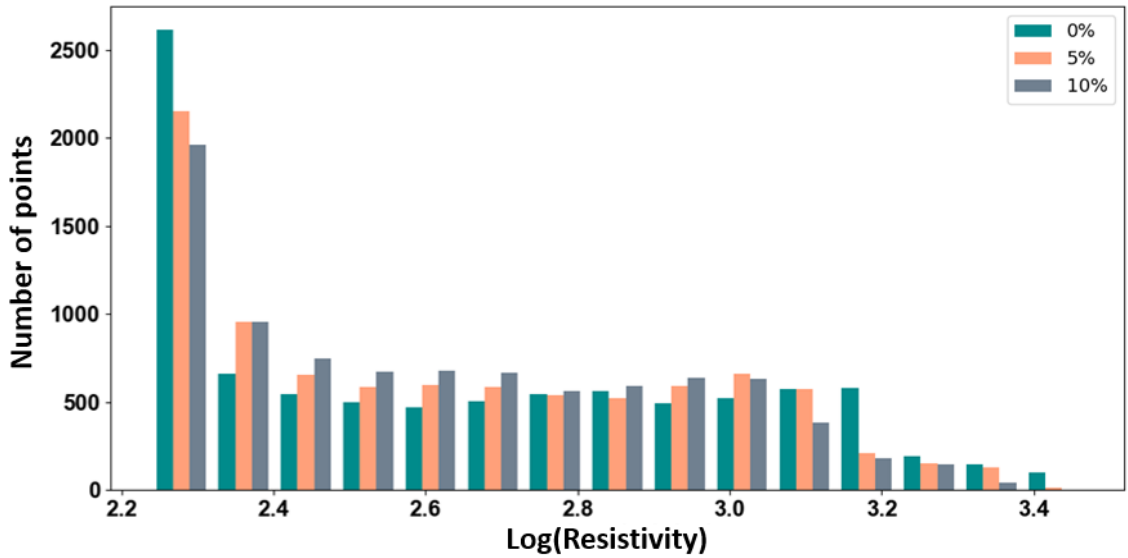

**Figure 10** Distribution of the logarithm of inverted electrical resistivity values in case B, for three noise values: 0%, 5% and 10%.

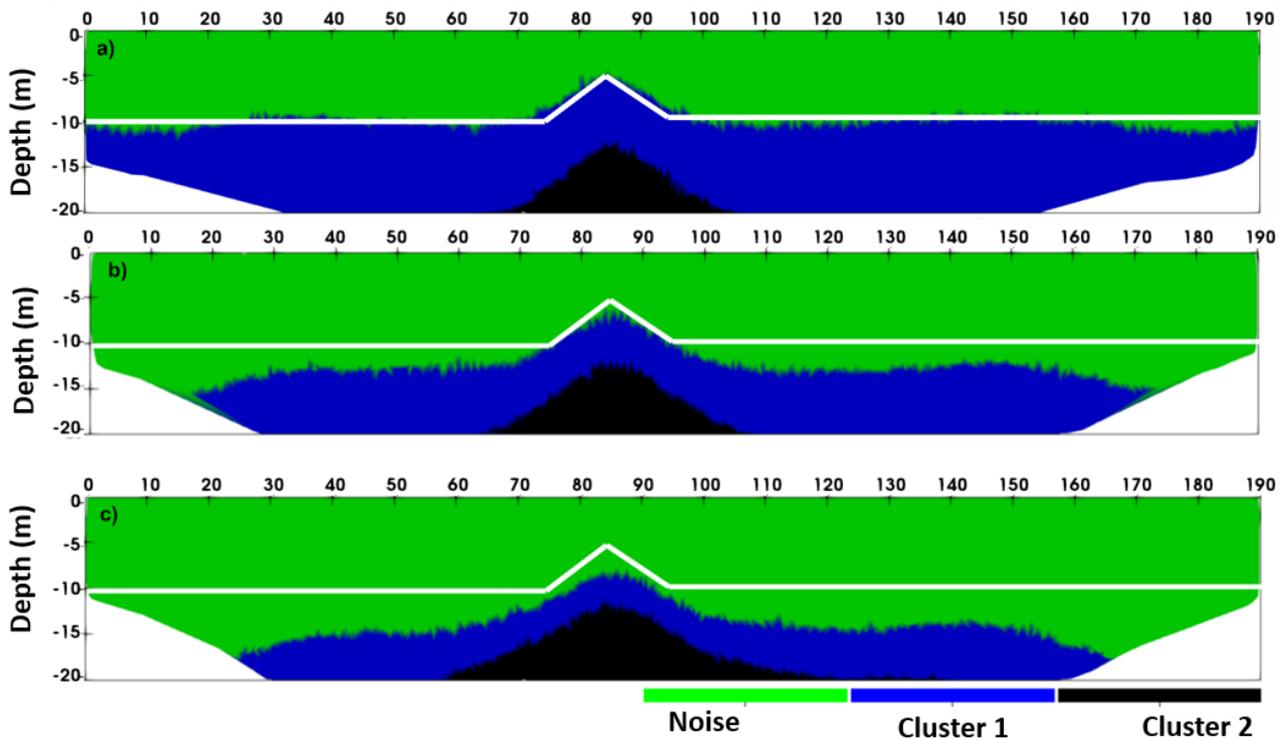

**Figure 11** Reconstructed structures obtained when the clustering algorithm is applied to case B, for noise values equal to (a) 0% (b) 5% and (c) 10%. The white line indicates the true position of the interface defined in the model.

## 4.3 Sensitivity to N

Another parameter that merits closer scrutiny is the number of points, *N*, since this is the first parameter to be chosen, and its choice impacts the remaining steps of the proposed methodology. To illustrate the influence of *N*, two different modelled structures were chosen. The first of these corresponds to case A, whereas the second one is provided by the tabular model with an anomaly in case C. In the previous analysis, the clustering results were represented by the distribution of inverted electrical resistivity values, which is not affected by changes in the value of *N*. Thus, in order to visualize the impact of *N* on the clustering, the distribution of the *N*-dist points has been plotted as shown in Figure 12b, for the first example.

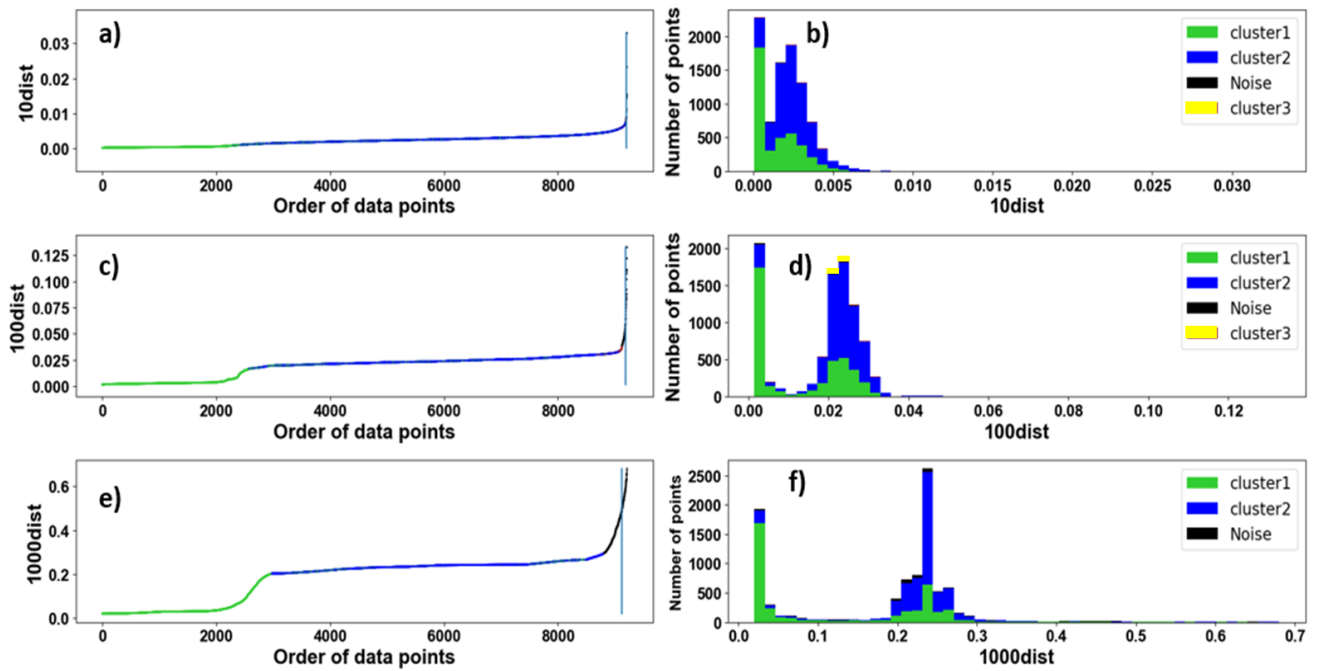

**Figure 12** Analysis of the impact of the number of points, *N*, on the density distribution represented by the *N*-dist plots (a-c-e), and by the distribution of the *N*-dist values (b-d-f). (a-b) *N* = 10. (c-d) *N* = 100. (e-f) *N* = 1000.

### 4.3.1 Case A: Two layers with a step

As the number of points *N* increases, *N*-dist increases because the algorithm calculates the distance to the *N* nearest points in terms of electrical resistivity. When *N* is high, the $N^{th}$ nearest neighbour may have a very different value of electrical resistivity, resulting in a higher number of *N*-dist data points, with an increasingly sparse distribution, as can be seen in Figure 12. Indeed, as N increases, *N*-dist increases, and its distribution is characterised by a clear transition from one layer to the next. Nevertheless, for all three cases, the algorithm was able to reconstruct the simulated structures (Figure *13*). Note that in the case of the two smallest values of *N*, i.e. *N* = 10 and *N* = 100, an additional cluster (yellow cluster 3 in Figure 13a and 13b) is produced, close to the noise data points. This occurs because a low value of *N* allows a cluster to be created within the noise data points. However, this type of cluster, in the vicinity of noise data points, and located at depths where the sensitivity is poor, should not be considered during the interpretation of the clustering results.

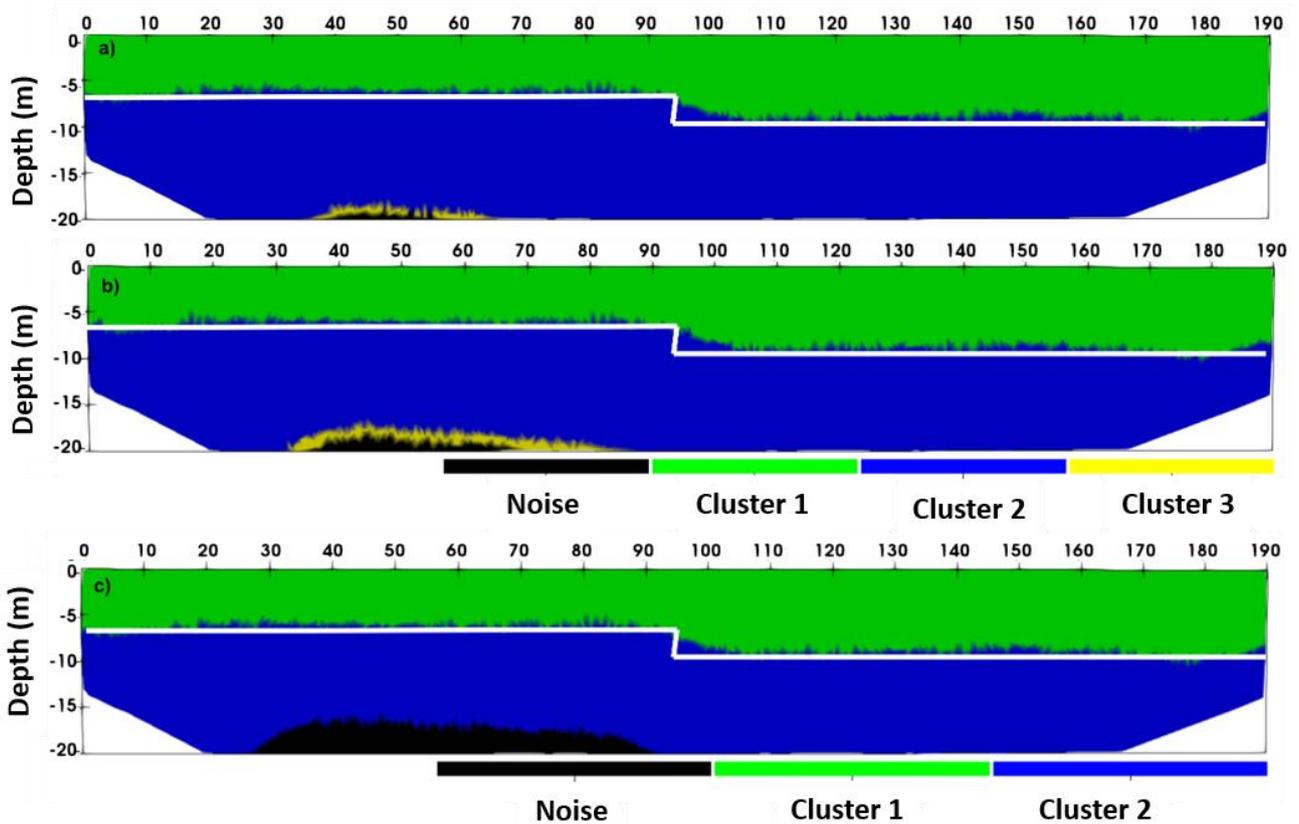

**Figure 13** Reconstructed profiles obtained by applying DBSCAN to the example (characterised by a step in the interface) of case A, using (a) *N* = 10, (b) *N* = 100 and (c) *N* = 1000. The white line indicates the true position of the interface between the two layers, as defined in the model.

### 4.3.2  Case C: Two layers and an anomaly

As in the case of the previous example, an increase in the number of points N for the example with an anomaly causes the distribution of *N*-dist to become more sparse, and also allows the transition between the two layers to become more visible in the *N*-dist distributions. However, unlike the case of the step model, the reconstructed structure is not always similar to the original shape of the simulated model (Figure 14). A low value for *N* tends to lower the reconstructed depth of the interface, and can also generate parasitic clusters around the noise points (Figure 14a). A low value for *N* allows data points to gather in clusters, even when they are sparsely distributed. Taking the example of the above values of *N*, although it would be possible to combine 10 points in a cluster, this would not be realistic for 1000 noisy data points. This is the reason for which the algorithm does not retrieve parasitic clusters when *N* is high. However, when the value of *N* is (too) high, the anomaly tends to be absorbed by the surrounding layer. As the anomaly corresponds to a small number of points (less than 1000), setting *N* = 1000 as the minimum number of points required to define a cluster causes the anomaly to merge with the first layer (Figure 14c). This analysis shows that *N* must be carefully chosen, depending on the expected structure of the terrain, and/or the extent of the anomalies it may contain.

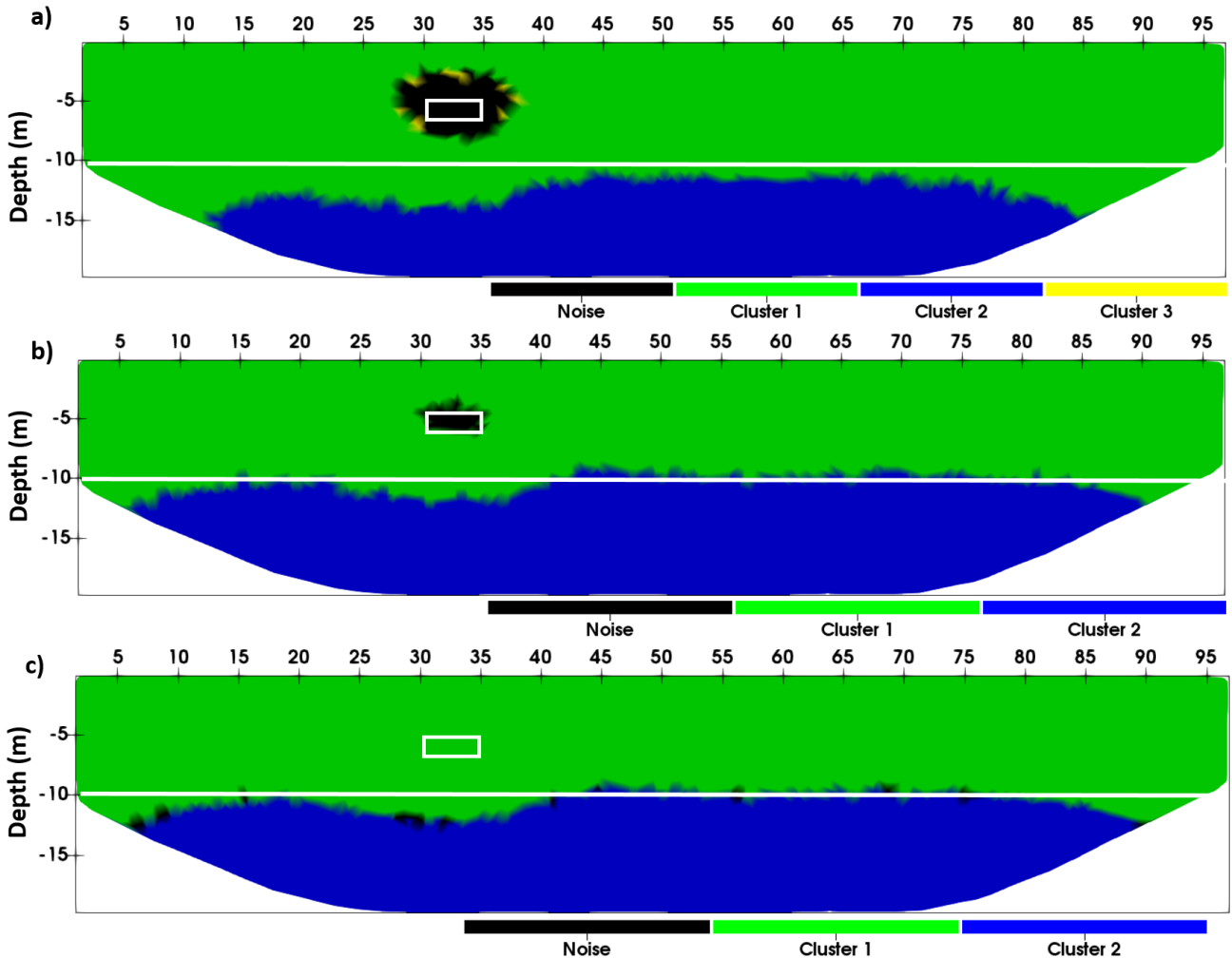

**Figure 14** Reconstructed profiles obtained by applying DBSCAN to the example of case C, using (a) *N* = 10, (b) *N* = 500 and (c) *N* = 1000. The white line indicates the true position of the interface between the two layers, as defined in the model.

## 5   Field case study

In order to study the effectiveness of our approach when it is applied to real data, we selected a set of ERT measurements recorded during the hydro-geophysical investigation of a shallow aquifer at the Orgeval basin, which is located 70 km east of Paris (Pasquet et al., 2015). The upper layers of this area are known to be strongly tabular. During this investigation, ERT measurements were recorded using a multi-channel resistivity meter, with a 96-electrode Wenner-Schlumberger array. The base electrode separation was 0.5 m. A geological log (Figure 15a) was used to describe the layers of soil. The shallow layer has a thickness of 0.25 m, corresponding to the agricultural soil. The second layer of table-land loess has a thickness of 3.75 m, and this covers a semi-infinite layer of Brie limestone.

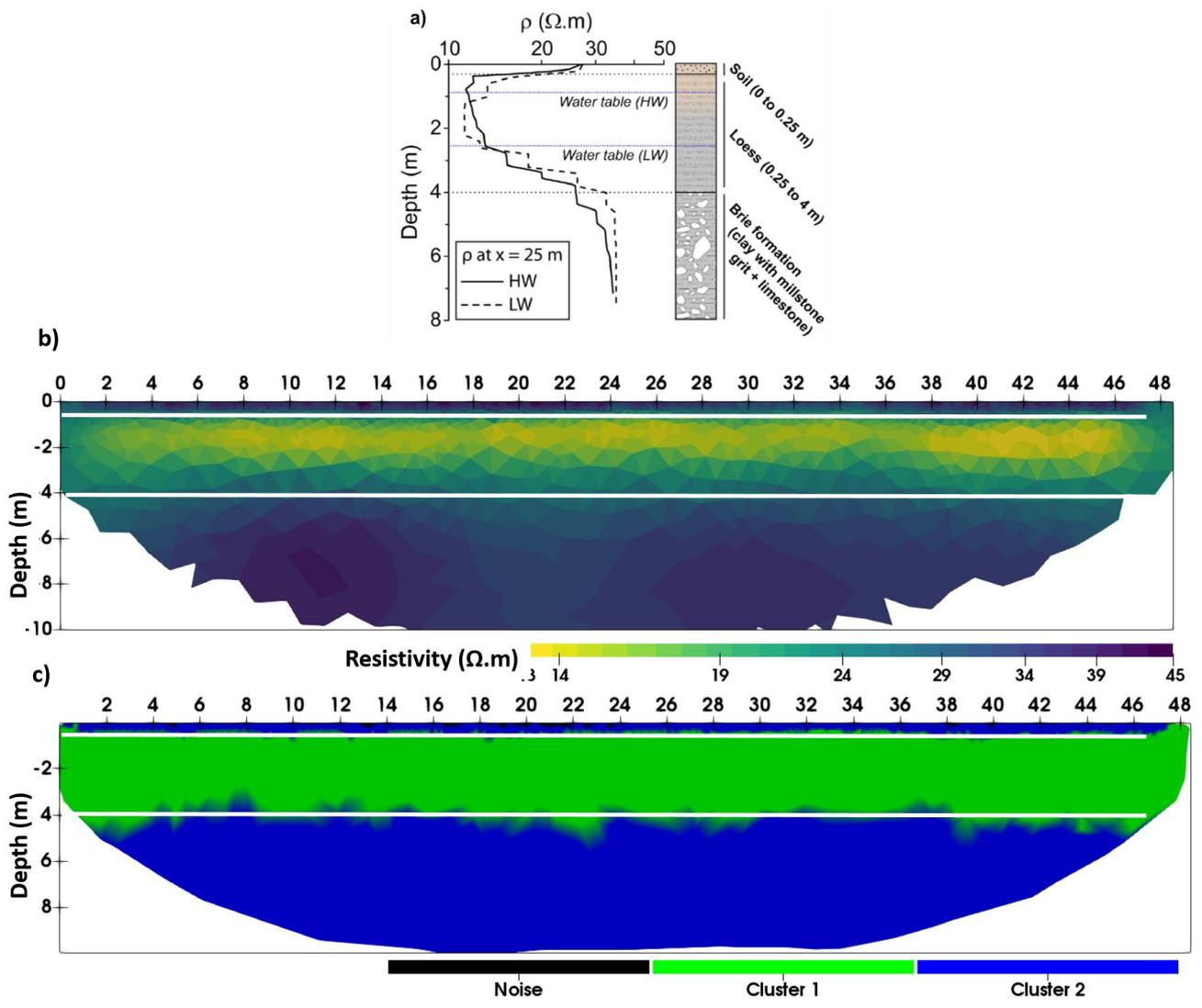

**Figure 15** Application of the DBSCAN clustering algorithm to field data measured at Orgeval. a) Interpreted geological log and electrical resistivity (Pasquet et al., 2015). b) Inverted electrical resistivity profile. c) Result obtained by applying the DBSCAN clustering method to the inverted electrical resistivity values in (b). The clustering is performed using N = 500 and $\varepsilon$ = 0.06. The white lines indicate the location of the layer interfaces, derived from the geological log.

Figure 15b presents the inverted electrical resistivity profile, showing the tabular, three-layer aspect of the prospected area. However, the resulting smoothed, inverted electrical profile does not allow the clear-cut physical interfaces between the layers to be identified. These interfaces are more accurately derived from the electrical resistivity log. By applying the clustering approach to the inverted electrical resistivity profile shown in Figure 15b, we retrieved the reconstructed soil structure shown in Figure 15c. The clustering produced by this analysis provides a detailed description of the site's stratigraphy. The first and second interfaces, at 0.25 m and 4 m respectively, are retrieved and found to be in good agreement with the geological log. No prior knowledge from the geological log was included in the clustering algorithm. However, the algorithm detects two clusters. The soil layer and the Brie formation are represented as merged into the same cluster, because they are characterized by the same value of electrical resistivity, i.e. from 30 Ωm to 35 Ωm. In order to discriminate between the two soil types in the clustering analysis, another geophysical method would be required, revealing different geophysical characteristics for each of these two layers.

## 6 Discussion and conclusions

In this research presents a proof of concept for the use of data mining techniques, DBSCAN in particular, for the improved, automated interpretation of geophysical sections. When applied to inverted electrical resistivity, it is able to identify clusters that are associated with distinct soil structures, and allows accurate reconstruction of the prospected area. This new approach to the interpretation of ERT profiles has been validated on simulated data. Using DBSCAN, we can recover sharp interfaces as well as the location and extent of anomalies, whilst avoiding the smoothness problem of previously reported inverted profiles. When applied to real field data, this technique provides a detailed description of the site stratigraphy, with an accurate determination of the position of layer interfaces. However, as can be seen in the real field example (Fig. 15), the algorithm assigns the same cluster to earth materials that are characterized by the same value of electrical resistivity, or the same geophysical parameter in general. This outcome is not surprising, since DBSCAN is an unsupervised algorithm, and the user is not involved with its interpretation. Users could use further judgement or analysis, such as geological sections, to distinguish between distinct unit(s) that could be represented by each cluster. The use of a different geophysical method more sensitive to the differences between adjacent, electrically similar materials/units, could also be considered if the aim of the study is not only to determine the soil structure but also to identify the different earth materials. An alternative approach would be to apply the clustering algorithm to two geophysical parameters at the same time (e.g., electrical resistivity and seismic velocity), as a way of coupling. The DBSCAN clustering analysis is performed quickly, and could therefore be used in an interpretation workflow, with negligible increase in overall computing time. This algorithm requires only two parameters, and does not require the user to have a deep understanding of data mining. In addition, it does not require any prior knowledge of the prospected area. The present study defines the steps needed to determine the two DBSCAN parameters. The approach described here could be applied to any other type of geophysical data that can be represented in the form of maps (e.g., magnetic, gravimetric), profiles (e.g., borehole logging), sections (e.g., seismic velocities), or images (e.g., thermal imaging). It is a useful approach for the study and characterization of discontinuities in different earth science applications (e.g. determination of saltwater/freshwater interfaces, detection of the interface between frozen and unfrozen areas, detection of cavities). This algorithm can also be directly applied to non-inverted data, such as apparent measurements. Although in the present study, DBSCAN was applied to 2D ERT profiles, the same approach could be applied to 2D apparent resistivity maps or 3D sections. In this case, the algorithm would also require the use of just two parameters, *N* and *ε*.

## 7 References

Amado, A., Cortez, P., Rita, P., & Moro, S., 2018. Research trends on Big Data in Marketing: A text mining and topic modeling based literature analysis. *European Research on Management and Business Economics*, **24**(1), 1–7.

Bauer, K., Moeck, I., Norden, B., Schulze, A., Weber, M., & Wirth, H., 2010. Tomographic P wave velocity and vertical velocity gradient structure across the geothermal site Groß Schönebeck (NE German Basin): Relationship to lithology, salt tectonics, and thermal regime. *Journal of Geophysical Research: Solid*

*Earth*, **115**(8).

Bièvre, G., Lacroix, P., Oxarango, L., Goutaland, D., Monnot, G., & Fargier, Y., 2017. Integration of geotechnical and geophysical techniques for the characterization of a small earth-filled canal dyke and the localization of water leakage. *Journal of Applied Geophysics*, **139**, 1–15.

Binley, A., 2015. Tools and techniques: DC Electrical Methods. *Treatise on Geophysics*, 233–259.

Borland, D., & Russell, M., 2007. Rainbow color map (Still) considered harmful. *IEEE Computer Society*, 14–17.

Calderon-Macias, C., K.Sen, M., & L.Stoffa, P., 2000. Artificial neural networks for parameter estimation in geophysics. *Geophysical Prospecting*, **48**, 21–47.

Cardarelli, E., & Fischanger, F., 2006. 2D data modelling by electrical resistivity tomography for complex subsurface geology. *Geophysical Prospecting*, **54**(2), 121–133.

Chambers, J. E., Wilkinson, P. B., Uhlemann, S., Sorensen, J. P. ., Roberts, C., Newell, A. J., et al., 2014. Derivation of lowland ripian wetland deposit architecture using geophysical image analysis and interface detection. *Water Resources Research*, **50**(7), 5886–5905.

Chen, Y., 2017. Automatic microseismic event picking via unsupervised machine learning. *Geophysical Journal International*, **212**(1), 88–102.

Dahlin, T., 2001. The development of electrical imaging techniques The development of DC resistivity imaging techniques. *Computers and Geosciences*, **27**, 1019–1029.

Day-Lewis, F. D., Singha, K., & Binley, A. M., 2005. Applying petrophysical models to radar travel time and electrical resistivity tomograms: Resolution-dependent limitations. *Journal of Geophysical Research: Solid Earth*, **110**(B8).

Dezert, T., Palma Lopes, S., Fargier, Y., & Côte, P., 2019. Geophysical and geotechnical data fusion for levee assessment - Interface detection with biased geophysical data. *24th European Meeting of Environmental and Engineering Geophysics*.

Ditmar, P. G., & Makris, J., 1996. Tomographic inversion of 2-D WARP data based on Tikhonov regularization. In *SEG Technical Program Expanded Abstracts 1996* (pp. 2015–2018). Society of Exploration Geophysicists.

Ellis, R. G., & Oldenburg, D. W., 1994. Applied geophysical inversion. *Geophysical Journal International*, **116**(1), 5–11.

Ester, M., Kriegel, H., Xu, X., & Sander, J., 1996. A density-based algorithm for discovering clusters in large spatial databases with noise. *Association for the Advancement of Artificial Intelligence*, **96**(34), 226–231.

Fargier, Y., Palma Lopes, S., Fauchard, C., François, D., & Côte, P., 2014. DC-Electrical Resistivity Imaging for embankment dike investigation: A 3D extended normalisation approach. *Journal of Applied Geophysics*, **103**, 245–256.

Finco, C., Pontoreau, C., Schamper, C., Massuel, S., Hovhannissian, G., & Rejiba, F., 2018. Time-domain electromagnetic imaging of a clayey confining bed in a brackish environment: A case study in the Kairouan Plain Aquifer (Kelbia salt lake, Tunisia). *Hydrological Processes*, **32**(26), 3954–3965.

Garambois, S., Voisin, C., Romero Guzman, M. A., Brito, D., Guillier, B., & Réfloch, A., 2019. Analysis of ballistic waves in seismic noise monitoring of water table variations in a water field site: added value from numerical modelling to data understanding. *Geophysical Journal International*, **219**(3), 1636–1647.

Goldman, M., Arad, A., Kafri, U., Gilad, D., & Melloul, A., 1989. Detection of fresh-water/sea-water interface by the time domain electromagnetic (TDEM) method in Israel. *Swim*, **70**, 329–344.

Han, J., Kamber, M., & Pei, J., 2011. *Data mining : concepts and techniques* (Elsevier). Morgan Kaufmann publishers.

Hsu, H. L., Yanites, B. J., Chen, C. chih, & Chen, Y. G., 2010. Bedrock detection using 2D electrical resistivity imaging along the Peikang River, central Taiwan. *Geomorphology*, **114**(3), 406–414.

Jin, G., Mendoza, K., Roy, B., & Buswell, D. G., 2019. Machine learning-based fracture-hit detection algorithm using LFDAS signal. *The Leading Edge*, **38**(7), 520–524.

Johansson, S., & Dahlin, T., 1996. Seepage monitoring in an earth embankment dam by repeated resistivity measurements. *European Journal of Engineering and Geophysics*, **1**, 229–247.

Jougnot, D., Jiménez-Martínez, J., Legendre, R., Le Borgne, T., Méheust, Y., & Linde, N., 2018. Impact of small-scale saline tracer heterogeneity on electrical resistivity monitoring in fully and partially saturated porous media: Insights from geoelectrical milli-fluidic experiments. *Advances in Water Resources*, **113**, 295–309.

Keller, G. V., & Frischknecht, F. C., 1966. Electrical methods in geophysical prospecting.

Krasnopolsky, V. M., & Schiller, H., 2003. Some neural network applications in environmental sciences. Part I: forward and inverse problems in geophysical remote measurements. *Neural Networks*, **16**(3–4), 321–334.

Kunetz, G., 1966. *Principles of Direct Current - Resistivity Prospecting*. Berlin: Gebrüder Borntraeger.

Ling, C., Revil, A., Abdulsamad, F., Qi, Y., Soueid Ahmed, A., Shi, P., et al., 2019. Leakage detection of water reservoirs using a Mise-à-la-Masse approach. *Journal of Hydrology*, **572**, 51–65.

Menke, W., 1989. *Geophysical data analysis: discrete inverse theory*. Academic Press.

Nagpal, A., Jatain, A., & Gaur, D., 2013. Review based on data clustering algorithms. *IEEE Conference on Information and Communication Technologies, ICT 2013*, 298–303.

Nicollo, M., 2014. Geophysical tutorial how to evaluate and compare color maps. *Society of Exploration Geophysicists*, 910–912.

De Pasquale, G., Linde, N., Doetsch, J., & Holbrook, W. S., 2019. Probabilistic inference of subsurface heterogeneity and interface geometry using geophysical data. *Geophysical Journal International*, **217**(2), 816–831.

Pasquet, S., Bodet, L., Longuevergne, L., Dhemaied, A., Camerlynck, C., Rejiba, F., & Guérin, R., 2015. 2D characterization of near-surface VP/VS : surface-wave dispersion inversion versus refraction tomography. *Near Surface Geophysics*, **13**(4), 315–331.

Rücker, C., Günther, T., & Spitzer, K., 2006. Three-dimensional modelling and inversion of DC resistivity data incorporating topography- II. Inversion. *Geophysical Journal International*, **166**(2), 495–505.

Rücker, C., Günther, T., & Wagner, F. M., 2017. pyGIMLi: An open-source library for modelling and inversion in geophysics. *Computers and Geosciences*, **109**, 106–123.

Russell, B., 2019. Machine learning and geophysical inversion — A numerical study. *The Leading Edge*, **38**(7), 512–519.

Samyn, K., Travelletti, J., Bitri, A., Grandjean, G., & Malet, J. P., 2012. Characterization of a landslide geometry using 3D seismic refraction traveltime tomography: The La Valette landslide case history. *Journal of Applied Geophysics*, **86**, 120–132.

Shirkhorshidi, A. S., Aghabozorgi, S., Wah, T. Y., & Herawan, T., 2014. Big data clustering: A review. *International Conference on Computational Science and Its Applications*, 707–720.

Simon, F., Koziol, A., & Thiesson, J., 2012. Investigating magnetic ghosts on an early middle age settlement: comparison of data from stripped and non-stripped areas. *Archaeological Prospection*, **145**, 142–145.

Soni, J., & Ansari, U., 2011. Predictive Data Mining for medical diagnosis heart disease prediction. *International Journal of Computer Applications*, **17**(8), 119–138.

Tan, A.-H., 1999. Text mining: The state of the art and the challenges. *Proceedings of the PAKDD 1999 Workshop on Knowledge Disocovery from Advanced Databases*, **8**, 65–70.

Tarantola, A., 2005. *Inverse problem theory and methods for model parameter estimation* (SIAM).

Thiesson, J., Rousselle, G., Simon, F. X., & Tabbagh, A., 2011. Slingram EMI prospection: Are vertical orientated devices a suitable solution in archaeological and pedological prospection? *Journal of Applied Geophysics*, **75**(4), 731–737.

Uhlemann, S., Hagedorn, S., Dashwood, B., Maurer, H., Gunn, D., Dijkstra, T., & Chambers, J., 2016. Landslide characterization using P- and S-wave seismic refraction tomography — The importance of elastic moduli. *Journal of Applied Geophysics*, **134**, 64–76.

Ward, W. O. C., Wilkinson, P. B., Chambers, J. E., Oxby, L. S., & Bai, L., 2014. Distribution-based fuzzy clustering of electrical resistivity tomography images for interface detection. *Geophysical Journal International*, **197**(1), 310–321.

Westergaard, D., Stærfeldt, H.-H., Tønsberg, C., Jensen, L. J., & Brunak, S., 2018. A comprehensive and quantitative comparison of text-mining in 15 million full-text articles versus their corresponding abstracts. *PLOS Computational Biology*, **14**(2), e1005962.

Xu, S., Sirieix, C., Riss, J., & Malaurent, P., 2017. A clustering approach applied to time-lapse ERT interpretation — Case study of Lascaux cave. *Journal of Applied Geophysics*, **144**, 115–124.

Zheng, Y., Zhang, Q., Yusifov, A., & Shi, Y., 2019. Applications of supervised deep learning for seismic interpretation and inversion. *The Leading Edge*, **38**(7), 526–533.